\documentclass[11pt]{article}
\usepackage{moriond,epsfig}

\def\be{\begin{equation}}
\def\ee{\end{equation}}
\def\bea{\begin{eqnarray}}
\def\eea{\end{eqnarray}}

\begin{document}
\vspace*{4cm}
\title{THE MSSM WITH DECOUPLED SCALARS AT THE LHC}

\author{ E. TURLAY\footnote{This work was done in collaboration with
    D. Zerwas (LAL-Orsay, Fr), N. Bernal and A. Djouadi (LPT-Orsay, Fr),
    M. Rauch and T. Plehn (U. of Edimburg, UK) and R. Lafaye
    (LAPP-Annecy, Fr)}}

\address{Laboratoire de l'Acc\'elerateur Lin\'eaire\\
Universit\'e Paris-Sud 11, 91898 Orsay cedex, France}

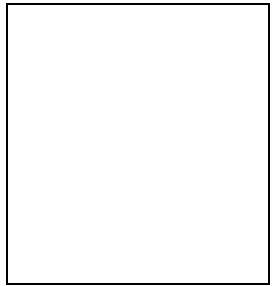
\maketitle
\abstracts{
  The discovery potential for the MSSM with heavy scalars at the LHC
  in the case of light inos is examined. We discuss the phenomenology
  of the model and the observables to determine the
  parameters. We show that for light gauginos, the model parameters
  can be constrained with a precision of the order of 15\%.
}

\section{Introduction}
Assuming a large soft--breaking scale for the MSSM
scalars~\cite{ArkaniHamed:2004fb,Giudice:2004tc,Wells:2004di,Kilian:2004uj,Bernal:2007uv}
pushes squarks, sfermions and heavy Higgses out of the kinematic reach of the
LHC without affecting the gaugino sector. The hierarchy problem will
not be solved without an additional logarithmic fine tuning
of the Higgs sector. Nevertheless, a model can be constructed to
provide a good candidate for dark matter and realize grand 
unification while minimizing proton decay and FCNCs. We investigate
the LHC phenomenology of the model, where all scalars are decoupled
from the low energy spectrum. We focus on gaugino--related signatures
to estimate the accuracy with which its underlying parameters can be
determined.

\section{Phenomenology}
The spectrum at the LHC is reduced to the gauginos, Higgsinos and the
light Higgs. At the intermediate scale $M_S$ the 
effective theory is matched to the full theory and the usual MSSM
renormalization group equations apply. The Higgsino mass parameter
$\mu$ and the ratio $\tan\beta$ in the Higgs sector correspond to
their MSSM counter parts. The gauginos masses $M_{1,2,3}$ and the
Higgs-sfermion-sfermion couplings unify, and $M_S$ replaces the
sfermion and the heavy Higgs' mass parameters.  This set resembles the
mSUGRA parameter set
except for $\tan\beta$ now playing the role of a matching parameter
(with the heavy Higgses being decoupled) rather than that of an actual
vev ratio~\cite{Drees:2005cp}.

We select our parameter point according to three constraints: first,
we minimize the amount of fine tuning necessary to bring the light
Higgs mass into the 100 to 200~GeV range and reduce $M_S$ to 10~TeV,
still well outside the LHC mass range.
The main reason for this low breaking scale is that we want the gluino
to decay inside the detector (preferably at the interaction point)
instead of being long--lived~\cite{Kraan:2004tz,Kilian:2004uj}.

Secondly, we obtain the correct relic dark--matter density $\Omega
h^2=0.111^{+0.006}_{-0.008}$~\cite{Spergel:2006hy}
by setting $\mu=290$ GeV and $M_2(M_{\rm GUT})=132.4$~GeV or $M_2(M_{\rm
  weak})=129$~GeV. This corresponds to the light--Higgs funnel
$m_{\rm LSP} \approx M_2/2 \approx M_h/2$, where the $s$-channel Higgs
exchange enhances the LSP annihilation rate. And finally, $m_h$ needs
to be well above the LEP limit, which we achieve by choosing
$\tan\beta=30$. We obtain $m_h = 129$~GeV,
$m_{\tilde{g}}=438$~GeV, chargino masses of 117 and 313~GeV, and
neutralino masses of 60, 117, 296, and 310~GeV
with a modified version of SuSpect~\cite{Bernal:2007uv,Djouadi:2002ze},
decoupling the heavy scalars from the MSSM RGEs.  $\tilde\chi^{0}_2$
and $\tilde\chi^{\pm}_1$ as well as $\tilde\chi^{0}_4$ and
$\tilde\chi^{\pm}_2$ are degenerate in mass. All supersymetric
particles and most notably the gluino are much lighter than in the
SPS1a parameter point
It is important to note that this feature is specific to our choice of
parameters and not generic in heavy--scalar models. As a consequence,
all LHC production cross sections are greatly enhanced with respect to
SPS1a.

\begin{table}[t]
\begin{center}
\begin{tabular}{c|c|c|c}
\hline
$\tilde g\tilde g$ & 68 pb & $\tilde\chi^{\pm}\tilde g$ & 311 fb\\
$\tilde\chi^{\pm}\tilde\chi^{0}$ & 12 pb & $\tilde\chi^{0}\tilde g$ & 223 fb\\
$\tilde\chi^{\pm}\tilde\chi^{\pm}$ & 6 pb  & $\tilde\chi^{0}\tilde\chi^{0}$ & 98 fb\\
\hline
\multicolumn{2}{c|}{Total} & \multicolumn{2}{c}{87 pb}\\
\end{tabular}
\caption{\label{splitsfitter_xsec} NLO cross sections for SUSY pair
  production at the LHC.}
\end{center}
\end{table}

Table~\ref{splitsfitter_xsec} shows the main (NLO) cross sections at
the LHC from
Prospino2~\cite{Beenakker:1996ch,Beenakker:1999xh,Plehn:1998nh}. The
SUSY production is dominated by gluino pairs whose rate is eight times
that of the SPS1a point: the lower gluino mass enlarges the available
phase space, while in addition the destructive interference between
$s$ and $t$--channel diagrams is absent. The second largest process is
the $\tilde\chi^{\pm}_1\tilde\chi^{0}_2$ production, which gives rise
to a 145~fb of hard-jet free, $e$ and $\mu$ trilepton signal, more
than a hundred times that of the SPS1a point.

\section{OBSERVABLES}
The first obvious observable is the light Higgs mass
$m_h$. Although slightly higher than in most MSSM points, $m_h$ can
still be measured in the Higgs to two photons decay
\cite{Bettinelli} ($m_h<150$ GeV). The systematic error on this
measurement is mainly due to the incertainty on the knowledge of the electromagnetic energy scale.

A measurement of the gluino pair production cross section appears feasible and could be
very helpful to determine $M_3$. The branching ratio of gluinos decaying through a
virtual squark into a chargino or a neutralino along with two
jets is 85\%. The chargino will in turn decay mostly into the LSP plus two
leptons or jets. Such events would feature at least 4 high-$p_T$
jets, a large amount of missing energy due to the two $\tilde\chi^{0}_1$ in the
final state and possibly leptons. The main backgrounds for such
signatures are $t\overline{t}$ pairs, $W+$jets and $Z+$jets
with respective production rates of 830 pb, 4640 pb and 220 pb
\cite{TDR:1999fr}. Despite these large cross sections, most of the
background can be eliminated by applying standard cuts on $\slash \hskip-3mmE_T$, the
number of high-$p_T$ jets as well as the effective
mass\footnote{$M_{\rm eff}=\slash\hskip-2mm E_T+\sum p_{T}(\rm jets)$.} which we
checked using a fast LHC-like simulation. The main source of
systematic errors for this observable is the 5\% error on the
knowledge of the luminosity. We take the theoretical error on the calculation
of the cross section to be roughly 20\%.

The next observable is the trilepton
signal. After gluino pairs, the next dominant channel is the direct
production of $\tilde\chi^{\pm}_1\tilde\chi^{0}_2$. 22\% of $\tilde\chi^{\pm}_1$s decay through a
virtual $W$ into an electron or muon and a neutrino and the
LSP. Similarly, 7\% of $\tilde\chi^{0}_2$s decay through a virtual $Z$
into an
Opposite-Sign-Same-Flavour lepton pair (OSSF) and the LSP. The
resulting signal features three leptons among which two are
OSSF, a large amount of missing transverse energy due to the two
LSPs plus the neutrino and no jet in the
hard process. The background for this signature is mainly $WZ$ and
$ZZ$ in which one of the leptons was non-identified or outside
acceptance. According to {\tt 
  PYTHIA} the lepton production ($e$ and $\mu$) rates are 386 fb
for $WZ$ and 73 fb for $ZZ$. The trilepton signal has a rate
of 145 fb, using {\tt SDECAY}~\cite{Muhlleitner:2004mka} for the
calculation of the branching ratios. Including identification efficiencies of 65\% for
electrons and of 80\% for muons~\cite{ADP:2007} gives rates of 
110 to 211 fb for the background and 40 to 74 fb for the signal before
any cut. A study with full detector simulation and reconstruction would
provide a better understanding of signal and background. As
in the previous case, the main 
source of systematic errors is the uncertainty on the luminosity. We
also take the theoretical error on the value of the trilepton cross
section to be roughly 20\%.

Within this trilepton signal lies another observable. 10\%
of $\tilde\chi^{0}_2$s decay into an OSSF lepton pair and the
LSP. The distribution of the invariant mass of the pair features a
kinematic upper edge whose value is
$m_{\tilde\chi^{0}_2}-m_{\tilde\chi^{0}_1}$. Such an observable gives
precious information on the neutralino sector and hence on $M_1$. The
systematic error is dominated by the lepton energy scale. The
statistical error was extracted from a {\tt ROOT} fit of the
$M_{\ell\ell}$ distribution and we estimate the theoretical accuracy
to be of the order of 1\%.

The last observable we use in this study is the ratio of gluino decays
including a $b$ quark to those not including a $b$. A systematic error
of 5\% due to the tagging of $b$-jets and a theoretical uncertainties
of 20\% are assumed.
\begin{table}[ht]
\begin{center}
\begin{tabular}{c|c|c|c|c|c}
\hline
\multicolumn{2}{c|}{Observables} & \multicolumn{2}{c|}{Exp. systematic
  errors} & Statistical errors & Theoretical\\
\hline
 & Value & Error & Source & 100 fb$^{-1}$ &  \\
\hline
$m_h$ & 128.8 GeV & 0.1\% & energy scale & 0.1\% &  4\%\\
$m_{\tilde\chi^{0}_2}-m_{\tilde\chi^{0}_1}$ & 57 GeV & 0.1\% & energy scale & 0.3\% &
 1\%\\
$\sigma(3\ell)$ & 145.2 fb & 5\% & luminosity & 3\% &  $20$\%\\
$R(\tilde g\rightarrow b/!b)$ & 0.11& 5\% & $b$-tagging & 0.3\%&  $20$\%\\
$\sigma(\tilde g\tilde g)$ & 68.2 pb & 5\% & luminosity & 0.1\%&  $20$\%\\
\hline
\end{tabular}
\caption{\label{obs}Summary of the observables and the corresponding errors.}
\end{center}
\end{table}

Table \ref{obs} summarises the value and error of the observables
assumed in this study. The third and fourth columns give the experimental
systematic errors and there source. The fifth column gives the statistical
errors for an integrated luminosity of 100 fb$^{-1}$ corresponding to
one year of data-taking at the LHC nominal luminosity. The last column
gives an estimation of the theoretical uncertainties.
\section{PARAMETER DETERMINATION}
We use different sets of errors for the fits. First we determine the
parameters in the
low statistic scenario ignoring theoretical uncertainties. Second we
assume an infinite statistic and therefore assume negligeable
statistical errors to estimate the ultimate precision barrier imposed
by experimental systematic errors. Finally the 
effect of theoretical uncertainties is estimated by including them into the
previous set. We expect these to dominate.\\
With no information on the squark and sfermion sector at all, except for
non-observation, we are forced to fix $M_S$ and $A_t$ and
set $M_2$ to be equal to $M_1$.
We fit the parameters to the observables using the {\tt Minuit} fitter.
The minimum of the $\chi^2$ is found by {\tt MIGRAD}. We start from a
point far from the nominal values ($\{M_1,M_3,\tan\beta,\mu\}=\{100,200,10,320\}$) and
reach the values reported in table \ref{fit}. 
 Errors are
determined with {\tt MINOS}.
Theoretical errors are treated as Gaussian.
\begin{table}[ht]
\begin{center}
\begin{tabular}{c|c|c|c|c|c|c|c|c}
\hline
Parameter & Nom. values & Fit values & \multicolumn{2}{|c}{Low stat.} & \multicolumn{2}{|c}{$\infty$
  stat.}&\multicolumn{2}{|c}{$\infty$ stat.$+$th}\\
\hline
$M_S$ & \multicolumn{2}{|c}{10 TeV} & \multicolumn{6}{|c}{fixed} \\
$A_t$ & \multicolumn{2}{|c}{0} & \multicolumn{6}{|c}{fixed} \\
$M_1$ & 132.4 GeV&132.8 GeV & \multicolumn{6}{|c}{$=M_2$} \\
\hline
$M_2$ & 132.4 GeV &132.8 GeV &6 &5\%&0.24&0.2\%&21.2&16\%\\
$M_3$ & 132.4 GeV & 132.7 GeV & 0.8 & 0.6\%&0.16&0.1\%&5.1&4\%\\
$\tan\beta$ & 30 & 28.3 & 60 & undet.&1.24&4\%&177&undet.\\
$\mu$ & 290 GeV & 288 GeV&3.8 & 1.3\%&1.1&0.4\%&48&17\%\\
\hline
\end{tabular}
\caption{\label{fit}Result of the fits. Errors on the determination of
the parameter are given for the three error sets. Both absolute and
relative values are given.}
\end{center}
\end{table}

Table \ref{fit} shows the result of the fits in both
absolute and relative values. It is interesting to note that
$\tan\beta$ in undetermined except in the case of infinite statistical
and theoretical accuracy. The quality of the trilepton
and gluino signals gives very good precision on the determination of
$M_1$ and $M_3$ even with low statisic. The inclusion
of theoretical uncertainties indeed decreases the accuracy but still
allows for a determination. $M_3$ only depends on the large
gluino signal and its decays, explaining its relative stability. $M_1$
and $M_2$ see the largest impact of theoretical errors.
This is because they depend on first order on the trilepton
cross-section and on second order on the $b$ to non $b$
gluino decays ratio both of which bear a large theoretical error.
\section{CONCLUSION}
The MSSM with heavy scalars can
very well satisfy current experimental and 
theoretical limits on physics beyond the standard model and also
solve a good number of issues present in the traditionnal MSSM. We
described its phenomenology at the LHC in the case of light inos and
showed that such a simple and light spectrum could lead to very high
production rates making the model discoverable. The main observable channels
are gluino pairs 
and the trilepton channel whose hard-jet free channel makes it well
distinct from SM and SUSY backgrounds. Other observables such as the
light Higgs mass, the $|m_{\tilde\chi^{0}_2}-m_{\tilde\chi^{0}_1}|$
kinematic edge and the $b$ to non $b$ producing gluino decays could
lead to a determination of most parameters to the level of a few
percent with 100 fb$^{-1}$ ignoring theoretical errors. In a
more realistic picture where we assumed non-zero theoretical errors,
we saw that most parameters can be determined with a precision of
15\%. We also saw that the scalar section including $\tan\beta$ could
only be poorly determined if at all.\\ 
New complementary
observables could help determine better the scalar sector. Equally, a
look at other parameter points will provide a more complete view of the discovery potential of
a MSSM with decoupled scalars at the LHC.


\section*{References}
\begin{thebibliography}{99}
\bibitem{ArkaniHamed:2004fb}N. Arkani-Hamed and S. Dimopoulos, {\em
  JHEP} {\bf 06} (2005) 073, {\tt[hep-th/0405159]}.
\bibitem{Giudice:2004tc}G.F. Giudice and A. Romanino, {\em Nucl. Phys.}
  {\bf B699} (2004) 64--89, {\tt[hep-ph/0406088]}
\bibitem{Wells:2004di}J.D. Wells, {\em Phys. Rev.}
  {\bf D71} (2005) 015013, {\tt[hep-ph/0411041]}.
\bibitem{Kilian:2004uj}W. Kilian, T.Plehn, P. Richardson and
  E. Schmidt, {\em Eur. Phys. J.}
  {\bf C39} (2005) 229--243, {\tt[hep-ph/0408088]}.
\bibitem{Bernal:2007uv}N. Bernal, A. Djouadi and P. Slavich, {\em JHEP}
  {\bf 07} (2007) 016, {\tt[arXiv:0705.1496 [hep-ph]]}.
\bibitem{Drees:2005cp}M. Drees {\tt[hep-ph/0501106]}.
\bibitem{Weinberg:1987dv}S. Weinberg, {\em Phys. Rev. Lett.} {\bf 59}
  (1987) 2607.
\bibitem{Kraan:2004tz}A. C. Kraan, {\em Eur. Phys. J.} {\bf C37}
  (2004) 91-104, {\tt hep-ex/0404001}.
\bibitem{Farrar:1978xj} G. R. Farrar and P. Fayet, {\em Phys. Lett.}
  {\bf B76} (1978) 575-579.
\bibitem{Spergel:2006hy} D. N. Spergel {\it et al.}, WMAP
  Collaboration, {\em
    Astrophys. J. Suppl.} {\bf 170} (2007) 377, {\tt
    astro-ph/0603449}.
\bibitem{Djouadi:2002ze} A. Djouadi, J.-L. Kneur and G. Moultaka, {\em
    Comput. Phys. Commun.} {\bf 176} (2007) 426-455, {\tt
    hep-ph/0211331}.
\bibitem{Beenakker:1996ch} W. Beenakker, R. Hopker, M. Spira and
  P. M. Zerwas, {\em Nucl. Phys} {\bf B492} (1997) 51-103, {\tt
    hep-ph/9610490}.
\bibitem{Beenakker:1999xh} W. Beenakker {\it et al.}, {\em
    Phys. Rev. Lett.} {\bf 83} (1999) 3780-3783, {\tt hep-ph/9906298}.
\bibitem{Plehn:1998nh} T. Plehn, (1998), {\tt hep-ph/9809319}.
\bibitem{Bettinelli} M. Bettinelli {\it et al.}, {\em
    ATL-PHYS-PUB-2007-013} (2007).
\bibitem{TDR:1999fr} ATLAS Collaboration, TDR vol. 2 CERN-LHCC-99-15
  (1999) 469.
\bibitem{Muhlleitner:2004mka} M. Muhlleitner, {\em Acta Phys. Polon.}
  {\bf B35} (2004) 2753-2766, {\tt hep-ph/0409200}.
\bibitem{ADP:2007} ATLAS Collaboration, Atlas Technical Paper.
\end{thebibliography}
\end{document}